# Institutional Shifts in Contribution to Indian Research Output during the last two decades

Vivek Kumar Singh[1,2,3], Mousumi Karmakar[4], Anurag Kanaujia[2]

[1]Department of Computer Science, University of Delhi, Delhi - 110007 (India)
[2]Delhi School of Analytics, University of Delhi, Delhi-110007 (India)
[3]Department of Computer Science, Banaras Hindu University, Varanasi - 221005 (India)
[4]Alliance College of Engineering and Design, Alliance University, Bengaluru-562106 (India)

**Abstract:** In the past few decades, India has emerged as a major knowledge producer, with research output being contributed by a diverse set of institutions ranging from centrally funded to state funded, and from public funded to private funded institutions. A significant change has been witnessed in Indian institutional actors during the last two decades, with various new private universities being set up and several new IITs, NITs, IISERs being established. Therefore, it is important to identify whether the composition of the list of the top 100 research output producing institutions of India has changed significantly during the recent two decades. This study attempted to analyse the changes during the two 10-year periods (2004-13 and 2014-23). The institutions which retain their position within top 100 during both periods are identified, along with the change in their positions. Similarly, institutions that were there in top 100 list during first time period (2004-13) and go out of top 100 list during second time period (2014-23) are also identified. In the same line, the new entrant institutions in the top 100 list during second time period (2014-23) are identified too. The results obtained indicate towards an institutional shift in the contribution to Indian research output.

## 1. Introduction

In the past few decades, India has emerged as a major knowledge producer, with some reports indicating that it ranks at third place in research output after the USA and China (National Science Board 2021, DST R&D Statistics Report 2021-22, Singh *et al.*, 2020). This research output is contributed by a diverse set of institutions ranging from centrally funded to state funded, and from public funded to private funded institutions. In fact, a recent study suggested that a majority of research publications in India are produced by the centrally funded institution systems (67.58%), which include, IITs, CSIR, Central Universities, Department of Atomic Energy (DAE) institutions, NITs, ICAR, DST etc. (Kanaujia *et al.*, 2022). In contrast, all other sources (state universities, private universities, Industry etc.) together account for the rest 32.42% of the publications (*ibid*). Some other studies have also identified the major institutions and systems contributing to Indian research output. For example, Prathap (2014) identified the role several elite institutions (also recognised as Institutes of National Importance) as contributors to Indian research output.

A significant change has been witnessed in Indian institutional actors during the last two decades. The CII-ICI report 2016 on research productivity of Indian universities underlined the increase in the number of private universities and a shift in their focus towards research activities as well. A look at the number of universities in India shows that the number of private universities has increased from a few private universities in 2003, to about 261 (in 2014) and about 430 (in 2023), state universities increased from 183 (in 2003) to 316 (in 2014) and 460 (in 2023), and Central Universities increased from 18 (in 2003), and 43 (in 2014) to 56 (in 2023) (AISHE 2014-15; 2021-22; University Grants Commission 2003; 2023). The number of Institutes of National Importance (INIs) also increased from 75 (in in 2014) to 153 (2022). Among these the most notable is the growth in number of private Universities and establishment of new INIs. Prathap & Sriram, (2017) while analysing the performance of selected private universities between 2013-2016, noted that while several of them had more than three times the number of faculty staff, and similar overall budget as that of IISc (the leading Institutes of National Importance in the period), their spending per faculty was much lower and the research activities were much less efficient resulting in lower publication output. However, a later study also noted that some of the private universities are attempting to position themselves more exclusively for research and innovation activities (Banshal et al., 2019).

The recently released NIRF rankings 2024 reflects a share of 23.48% publications by the private institutions/universities, and a share of 6% by the State Private Universities, in the national research output specifically from top 100 ranked institutions in NIRF (Department of Higher Education, 2024). These changes suggest towards a possible shift in the direction of the academic ecosystem where new universities/institutions may assume the centre stage in terms of their output in the research publications. Some other recent studies have also indicated that the rate of publication output of some of the major institutions have slowed down in the past few years, and several new research institutions have come to the fore in the R&D landscape (Kanaujia *et al.*, 2022; 2023). Thus, it becomes important to look at the changes in the list of institutions among the top contributors to Indian research. This study attempts to address this gap by identifying the major changes in the composition of the list of the top 100 research output producing institutions during the recent two decades, 2004-13 and 2014-23. The institutions which retain their position within top 100 during both periods are identified, along with the change in their positions. Similarly, institutions that were there in top 100 list during first time period (2004-13) and go out of top 100 during second time period (2014-23) are also identified. New entrant institutions in the top 100 list during second time period (2014-23) are identified too. Thus, this work attempts to answer following research questions:

**RQ1:** Which Indian institutions figure in the list of top 100 research output producing institutions during the two recent decades (2004-13 and 2014-23)?

**RQ2:** What are the major changes in the composition of the top 100 list during 2004-13 and 2014-23, and what do these changes indicate?

## 2. Related Work

Over the past few years, performance assessment and ranking of Indian institutions has become an area of policy and academic interest at national level. The institutional rankings promoted by the Ministry of Education, Govt. of India, namely, National Institutional Ranking Framework (NIRF) and Atal Ranking of Institutions on Innovation Achievements (ARIIA) can be seen as two examples (MHRD, 2017). In addition to these efforts, various researchers have also analysed the research performance of an individual or a group of institutions.

The research performance of India and Indian Institutions has been compared to institutions in different countries on criteria of research productivity (Prathap 2017; Prathap & Sriram, 2017; Prathap 2018; Singh et al., 2020). At national level, several individual studies have been conducted to analyse the performances of institutions or institution systems, such as the AIIMS

(Nishavathi & Jeyshankar, 2018), IITs (Prathap & Gupta, 2009; Prathap, 2013; Banshal *et al.*, 2017), NITs (Bala & Kumari, 2013; Banshal, Solanki & Singh, 2018), IIMs (Tyagi, 2022; Singh, Nandy & Singh, 2022), Indian Medical Institutions (Ray et al., 2016), IISERs (Solanki, Banshal & Singh, 2016), ICARs (Suresh & Thanuskodi, 2019), Central Universities (Basu *et al.*, 2016; Marisha, Banshal & Singh 2017), Private universities (Prathap & Sriram, 2017; Banshal, Singh & Mayr, 2019) and research intensive higher education institutions (Nishy *et al.*, 2012; Prathap 2014). More recently, a group of researchers developed a portal named 'Indian Science Reports' (www.indianscience.net) that benchmarks research output related indices and analysis for 1000 Indian institutions (Singh et al., 2022).

Though there have been several previous studies on analysis of research output and performance of Indian institutions, no previous study tried to see the institutional shifts in the contribution to Indian research output over time. It is not known as to whether and how the composition of the list of the top 100 research output producing institutions have changed during last 20 years. This study attempts to address this gap by identifying the major changes in the composition of the list of the top 100 research output producing institutions during the recent two decades, 2004-13 and 2014-23. The institutions that gain rank in the list, those that lose and go out of the list, and those that are able to retain their position, are identified. A discussion on the type and nature of the institutions losing and gaining in the list is developed.

## 3. Data & Method

The study has obtained research publication data (restricted to document types 'article' and 'review') of the Indian institutions from the Web of Science (WoS) database by using the following search query.

*CU="India" AND DT="Article" OR "Review" AND PY="2004-2023"*

The database search resulted in a total of 12,74,094 publication records. The data so obtained was divided into two groups corresponding to the two time periods, 2004-13 and 2014-23. For both the time periods, a list of top institutions in terms of research output is identified. The union of lists for the two time periods, 2004–2013 and 2014–2023, is also obtained. A total of 140 individual institutions figure in the union list.

To analyse the different trends and indicators, programs written in Python were used. The different indicators computed for the given list of institutions include total research publications, proportionate share to Indian research output, and the compounded annual growth rate (CAGR) of publications in the two analysis periods.

The changes in these indicators for the different institutions were observed and have been reported in the results section. The institutions which entered and those which slipped out of the top 100 list in the 2014-2023 period were identified and reported. A discussion around the type and nature of these institution is developed.

## 4. Results

### *Research output of top 100 Indian institutions during the two time periods of 10-year each*

For the top 100 institutions in the 2004-2013 period, the total publication count was 2,23,532 publications, among them which IISc Bangalore had the highest number of publications i.e., 13,965 followed by IIT Kharagpur (10,561) and University of Delhi (9,456). In the 2014-2023 period, the total publication count of the top 100 institutions was 4,60,813 publications (Table 1). Among them the top three institutions remained in their positions with 21976, 21349 and

21109 publications, respectively. In the top 10 institutions, IIT Kanpur moved from 9th position to 14th position, Jadavpur University moved from 10th position to 22nd position, IIT Roorkee moved from 12th to 10th position, and Vellore Institute of Technology moved from 45th to 8th position. The total number of publications for the institute at 100th position in 2004-2013 period was 1107 while the same for 2014-2023 period was 3407, which coincides with the overall increase in publication output from India. In table 1, the positions of institutes in both periods are provided and arrows are used to indicate change in the position. Movements within top 100 are shown by single arrows (↑ or ↓), and in or out of top 100 is shown by double arrows (⇑ or ⇓), while dash (-) shows no change in position.

Notably, 40 institutions in the top 100 of 2004-2013 slipped out of the top 100 in the 2014-2023 period (Table 1). The overall CAGR (2004-2023) was highest for Saveetha Institute of Medical Technical Science at 53.29% and it moved from 420[th] position with 198 publications during 2004-2013 to 20th position with 10,777 publications during 2014-2023 in the top 100 institutions. It may also be noted from the table that proportionate share of research contribution of almost all the institutions have decreased from first to second time period.

**Table 1: Research Output of top 100 Indian Institutions during 2004-2013 and 2014-2023**
**(Union of two sets)**

| S. No. | Name of the Institution | 2004-2013 | | | | 2014-2023 | | | | CAGR3 (2004-2023) |
|---|---|---|---|---|---|---|---|---|---|---|
| | | TP | Rank | CAGR1 (2004-2013) | % share | TP | Rank | CAGR2 (2014-2023) | % share | |
| 1 | Indian Institute of Science (IISc) Bangalore | 13,965 | 1 | 6.58 | 3.15 | 21,976 | 1 (-) | 3.32 | 1.82 | 4.99 |
| 2 | Indian Institute of Technology (IIT) Kharagpur | 10,561 | 2 | 9.23 | 2.38 | 21,349 | 2 (-) | 7.48 | 1.77 | 8.43 |
| 3 | University of Delhi | 9,456 | 3 | 12.36 | 2.13 | 21,109 | 3 (-) | 8.21 | 1.75 | 10.51 |
| 4 | Indian Institute of Technology (IIT) Delhi | 8,446 | 4 | 8.3 | 1.9 | 19,957 | 5 (↓) | 10.77 | 1.66 | 9.65 |
| 5 | Indian Institute of Technology (IIT) Madras | 8,231 | 5 | 10.92 | 1.86 | 20,465 | 4 (↑) | 10.50 | 1.7 | 10.62 |
| 6 | Banaras Hindu University (BHU) | 8,047 | 6 | 15.11 | 1.81 | 17,360 | 7 (↓) | 8.16 | 1.44 | 11.13 |
| 7 | Indian Institute of Technology (IIT) Bombay | 7,848 | 7 | 9.52 | 1.77 | 19,551 | 6 (↑) | 8.97 | 1.62 | 9.28 |
| 8 | All India Institute of Medical Sciences (AIIMS) New Delhi | 7,285 | 8 | 9.57 | 1.64 | 16,273 | 9 (↓) | 8.50 | 1.35 | 9.18 |
| 9 | Indian Institute of Technology (IIT) Kanpur | 7,104 | 9 | 6.49 | 1.6 | 13,316 | 14 (↓) | 7.70 | 1.1 | 7.19 |
| 10 | Jadavpur University | 6,742 | 10 | 10.67 | 1.52 | 12,817 | 15 (↓) | 4.47 | 1.06 | 7.81 |
| 11 | Tata Institute of Fundamental Research (TIFR) | 6,404 | 11 | 6.44 | 1.44 | 10,564 | 22 (↓) | 2.38 | 0.88 | 4.13 |
| 12 | Indian Institute of Technology (IIT) Roorkee | 5,870 | 12 | 16.21 | 1.32 | 15,997 | 10 (↑) | 10.17 | 1.33 | 13.38 |
| 13 | CSIR-Indian Institute of Chemical Technology (IICT) | 5,427 | 13 | 3.83 | 1.22 | 6,070 | 38 (↓) | -4.04 | 0.5 | 1.39 |
| 14 | Anna University | 5,322 | 14 | 14.13 | 1.19 | 13,725 | 13 (↑) | 5.19 | 1.14 | 9.98 |
| 15 | Panjab University | 4,945 | 15 | 10.9 | 1.11 | 10,696 | 21 (↓) | 7.44 | 0.89 | 9.01 |
| 16 | Aligarh Muslim University | 4,875 | 16 | 14.12 | 1.1 | 11,244 | 19 (↓) | 9.60 | 0.93 | 11.73 |
| 17 | University of Calcutta | 4,642 | 17 | 14.42 | 1.05 | 9,793 | 24 (↓) | 2.88 | 0.81 | 8.74 |
| 18 | Post Graduate Institute of Medical Education Research (PGIMER) Chandigarh | 4,469 | 18 | 12.02 | 1.01 | 11,898 | 18 (-) | 8.89 | 0.99 | 10.83 |

| | | | | | | | | | |
|---|---|---|---|---|---|---|---|---|---|
| 19 | CSIR-National Chemical Laboratory (NCL) | 4,089 | 19 | 1.47 | 0.92 | 5,137 | 46 (↓) | -2.88 | 0.43 | 0.44 |
| 20 | Indian Association for the Cultivation of Science (IACS) Jadavpur | 4,005 | 20 | 6.31 | 0.9 | 4,540 | 60 (↓) | -0.34 | 0.38 | 2.99 |
| 21 | Annamalai University | 3,932 | 21 | 13.54 | 0.89 | 5,549 | 42 (↓) | 1.65 | 0.46 | 7.94 |
| 22 | University of Madras | 3,531 | 22 | 6.49 | 0.8 | 4,831 | 52 (↓) | 5.11 | 0.4 | 5.36 |
| 23 | Indian Institute of Technology (IIT) Guwahati | 3,517 | 23 | 23.5 | 0.79 | 12,743 | 16 (↑) | 11.06 | 1.06 | 17.19 |
| 24 | University of Hyderabad | 3,501 | 24 | 12.27 | 0.79 | 7,127 | 32 (↓) | 5.44 | 0.59 | 8.85 |
| 25 | Manipal Academy of Higher Education (MAHE) | 3,369 | 25 | 25.59 | 0.76 | 14,892 | 12 (↑) | 17.63 | 1.24 | 20.47 |
| 26 | Jawaharlal Nehru University New Delhi | 2,901 | 26 | 11.11 | 0.65 | 8,539 | 28 (↓) | 8.66 | 0.71 | 10.61 |
| 27 | CSIR-National Physical Laboratory (NPL) | 2,725 | 27 | 12.21 | 0.61 | 3,694 | 87 (↓) | 0.24 | 0.31 | 6.80 |
| 28 | Savitribai Phule Pune University | 2,692 | 28 | 12.42 | 0.61 | 6,444 | 36 (↓) | 8.84 | 0.53 | 10.61 |
| 29 | University of Mysore | 2,684 | 29 | 9.8 | 0.6 | 3,105 | 113 (⇊) | 2.64 | 0.26 | 3.57 |
| 30 | Saha Institute of Nuclear Physics | 2,640 | 30 | 9.14 | 0.6 | 4,330 | 68 (↓) | -2.83 | 0.36 | 3.38 |
| 31 | Indira Gandhi Centre for Atomic Research (IGCAR) | 2,621 | 31 | 11.34 | 0.59 | 3,946 | 76 (↓) | -1.22 | 0.33 | 4.31 |
| 32 | Indian Statistical Institute Kolkata | 2,492 | 32 | 7.09 | 0.56 | 3,905 | 79 (↓) | 5.53 | 0.32 | 5.23 |
| 33 | CSIR-Central Drug Research Institute (CDRI) | 2,481 | 33 | 7.02 | 0.56 | 3,218 | 107 (⇊) | -0.98 | 0.27 | 3.81 |
| 34 | Jawaharlal Nehru Center for Advanced Scientific Research (JNCASR) | 2,418 | 34 | 5.98 | 0.54 | 3,132 | 110 (⇊) | -4.34 | 0.26 | 1.75 |
| 35 | University of Rajasthan | 2,385 | 35 | 4.16 | 0.54 | 2,032 | 191 (⇊) | 2.39 | 0.17 | 1.03 |
| 36 | Guru Nanak Dev University | 2,308 | 36 | 10.62 | 0.52 | 4,801 | 54 (↓) | 5.96 | 0.4 | 7.94 |
| 37 | ICAR-Indian Veterinary Research Institute | 2,288 | 37 | 8.01 | 0.52 | 3,581 | 92 (↓) | 0.03 | 0.3 | 4.59 |
| 38 | Sanjay Gandhi Postgraduate Institute of Medical Sciences | 2,267 | 38 | 10.03 | 0.51 | 3,744 | 84 (↓) | 5.82 | 0.31 | 7.35 |
| 39 | Jamia Millia Islamia | 2,195 | 39 | 15.09 | 0.49 | 7,537 | 30 (↑) | 14.93 | 0.63 | 14.63 |
| 40 | Indian Institute of Technology BHU Varanasi (IIT-BHU) Varanasi | 2,175 | 40 | 16.9 | 0.49 | 8,709 | 26 (↑) | 17.78 | 0.72 | 16.50 |
| 41 | University of Allahabad | 2,162 | 41 | 15.52 | 0.49 | 3,277 | 104 (⇊) | 7.43 | 0.27 | 10.21 |
| 42 | Institute of Chemical Technology Mumbai | 2,102 | 42 | 10.46 | 0.47 | 5,052 | 47 (↓) | 7.51 | 0.42 | 9.12 |
| 43 | Cochin University Science Technology | 2,088 | 43 | 9.08 | 0.47 | 3,946 | 75 (↓) | 10.59 | 0.33 | 9.14 |
| 44 | Sri Venkateswara University | 2,069 | 44 | 11.25 | 0.47 | 2,559 | 142 (⇊) | -9.15 | 0.21 | -0.31 |
| 45 | Vellore Institute of Technology (VIT) | 2,065 | 45 | 61.02 | 0.46 | 17,272 | 8 (↑) | 21.93 | 1.43 | 39.75 |
| 46 | Christian Medical College Hospital (CMCH) Vellore | 2,041 | 46 | 12.6 | 0.46 | 4,971 | 48 (↓) | 5.84 | 0.41 | 9.71 |
| 47 | Maharaja Sayajirao University Baroda | 2,014 | 47 | 10.39 | 0.45 | 3,117 | 111 (⇊) | 2.98 | 0.26 | 5.29 |
| 48 | Mangalore University | 1,979 | 48 | 6.62 | 0.45 | 2,015 | 193 (⇊) | 6.78 | 0.17 | 6.29 |
| 49 | Andhra University | 1,977 | 49 | 10.43 | 0.45 | 3,228 | 106 (⇊) | 4.25 | 0.27 | 5.19 |
| 50 | Bharathidasan University | 1,935 | 50 | 11.21 | 0.44 | 5,158 | 44 (↑) | 10.83 | 0.43 | 11.23 |
| 51 | Jamia Hamdard University | 1,868 | 51 | 17.34 | 0.42 | 4,483 | 62 (↓) | 8.96 | 0.37 | 12.31 |
| 52 | CSIR-Central Food Technological Research Institute (CFTRI) | 1,855 | 53 | 4.4 | 0.42 | 2,158 | 176 (⇊) | -1.29 | 0.18 | 1.71 |
| 53 | CSIR-National Institute Interdisciplinary Science Technology (NIIST) | 1,855 | 52 | 4.06 | 0.42 | 2,586 | 139 (⇊) | 1.76 | 0.22 | 3.61 |

| | | | | | | | | | |
|---|---|---|---|---|---|---|---|---|---|
| 54 | National Institute Of Technology (NIT) Tiruchirappalli | 1,852 | 54 | 25.88 | 0.42 | 6,699 | 34 (↑) | 13.03 | 0.56 | 18.30 |
| 55 | Shivaji University | 1,849 | 55 | 15.5 | 0.42 | 2,927 | 120 (⇊) | 4.07 | 0.24 | 8.35 |
| 56 | Bharathiar University | 1,747 | 56 | 13.06 | 0.39 | 5,993 | 39 (↑) | 5.61 | 0.5 | 9.84 |
| 57 | University of Mumbai | 1,706 | 57 | -4.5 | 0.38 | 2,188 | 172 (⇊) | 12.32 | 0.18 | 2.15 |
| 58 | Punjab Agricultural University | 1,701 | 58 | 6.09 | 0.38 | 4,849 | 51 (↑) | 14.36 | 0.4 | 10.87 |
| 59 | CSIR-Central Leather Research Institute (CLRI) | 1,700 | 59 | 5.54 | 0.38 | 2,114 | 185 (⇊) | -2.41 | 0.18 | 1.74 |
| 60 | Birla Institute of Technology Science Pilani (BITS) Pilani | 1,678 | 60 | 18.41 | 0.38 | 9,320 | 25 (↑) | 17.84 | 0.77 | 18.81 |
| 61 | Lucknow University | 1,639 | 61 | 9.36 | 0.37 | 2,881 | 123 (⇊) | 7.26 | 0.24 | 7.36 |
| 62 | Madurai Kamaraj University | 1,617 | 62 | 9 | 0.36 | 2,707 | 129 (⇊) | 1.25 | 0.23 | 4.92 |
| 63 | Indian Institute of Engineering Science Technology Shibpur (IIEST) | 1,611 | 63 | 13.86 | 0.36 | 4,688 | 58 (↑) | 7.28 | 0.39 | 10.37 |
| 64 | Visva Bharati University | 1,609 | 64 | 21.23 | 0.36 | 4,231 | 71 (↓) | 4.50 | 0.35 | 12.41 |
| 65 | Kalyani University | 1,559 | 65 | 13.57 | 0.35 | 3,034 | 117 (⇊) | 2.48 | 0.25 | 6.80 |
| 66 | University of Burdwan | 1,558 | 66 | 12.2 | 0.35 | 3,423 | 99 (↓) | 6.49 | 0.28 | 8.15 |
| 67 | Tata Memorial Centre (TMC) | 1,551 | 67 | 15.26 | 0.35 | 4,656 | 59 (↑) | 8.46 | 0.39 | 13.23 |
| 68 | Karnatak University | 1,535 | 68 | 3.46 | 0.35 | 2,125 | 179 (⇊) | 4.81 | 0.18 | 3.41 |
| 69 | King George S Medical University | 1,517 | 69 | 22.17 | 0.34 | 3,603 | 91 (↓) | 3.69 | 0.30 | 9.07 |
| 70 | CSIR-Indian Institute of Chemical Biology (IICB) | 1,512 | 70 | 8.06 | 0.34 | 2,005 | 194 (⇊) | -2.89 | 0.17 | 2.74 |
| 71 | Thapar Institute of Engineering Technology | 1,494 | 71 | 26.39 | 0.34 | 7,896 | 29 (↑) | 12.69 | 0.65 | 19.29 |
| 72 | Punjabi University | 1,484 | 72 | 29.91 | 0.33 | 3,562 | 93 (↓) | 4.96 | 0.3 | 16.29 |
| 73 | CSIR-National Institute of Oceanography (NIO) | 1,476 | 73 | 12.32 | 0.33 | 2,212 | 168 (⇊) | 0.45 | 0.18 | 6.85 |
| 74 | CSIR-Central Electrochemical Research Institute (CECRI) | 1,475 | 76 | 7.59 | 0.33 | 2,593 | 138 (⇊) | 0.92 | 0.22 | 5.49 |
| 75 | Kurukshetra University | 1,475 | 74 | 14.63 | 0.33 | 2,432 | 149 (⇊) | 11.08 | 0.20 | 9.50 |
| 76 | Physical Research Laboratory India | 1,475 | 75 | 5.11 | 0.33 | 2,631 | 134 (⇊) | 4.91 | 0.22 | 5.53 |
| 77 | Bangalore University | 1,472 | 77 | 7.22 | 0.33 | 1,752 | 218 (⇊) | -2.17 | 0.15 | 1.30 |
| 78 | Jawaharlal Nehru Technological University Hyderabad | 1,460 | 78 | 31.39 | 0.33 | 2,475 | 146 (⇊) | -3.46 | 0.21 | 10.24 |
| 79 | Inter University Accelerator Centre | 1,458 | 79 | 7.43 | 0.33 | 1,888 | 203 (⇊) | 2.46 | 0.16 | 4.81 |
| 80 | National Institute of Mental Health Neurosciences India | 1,435 | 80 | 16.8 | 0.32 | 4,402 | 66 (↑) | 7.40 | 0.37 | 11.84 |
| 81 | UGC DAE Consortium For Scientific Research | 1,433 | 81 | 8.01 | 0.32 | 2,716 | 128 (⇊) | 4.96 | 0.23 | 7.12 |
| 82 | CSIR-Central Salt Marine Chemical Research Institute (CSMCRI) | 1,428 | 82 | 13.68 | 0.32 | 2,333 | 156 (⇊) | 4.03 | 0.19 | 8.56 |
| 83 | National Institute of Technology Rourkela | 1,417 | 83 | 33.44 | 0.32 | 8,556 | 27 (↑) | 15.77 | 0.71 | 24.17 |
| 84 | SN Bose National Centre for Basic Science (SNBNCBS) | 1,393 | 84 | 9.16 | 0.31 | 2,257 | 163 (⇊) | 3.76 | 0.19 | 6.36 |
| 85 | Tata Memorial Hospital | 1,389 | 85 | 14.48 | 0.31 | 3,703 | 86 (↓) | 7.00 | 0.31 | 11.86 |
| 86 | CSIR-Centre for Cellular Molecular Biology (CCMB) | 1,386 | 86 | 6.87 | 0.31 | 1,494 | 255 (⇊) | -1.88 | 0.12 | 1.09 |
| 87 | Raja Ramanna Centre for Advanced Technology | 1,361 | 87 | 9.02 | 0.31 | 1,959 | 196 (⇊) | -1.64 | 0.16 | 4.39 |

| | | | | | | | | | |
|---|---|---|---|---|---|---|---|---|---|
| 88 | CSIR-National Geophysical Research Institute (NGRI) | 1,352 | 88 | 9.22 | 0.3 | 1,860 | 206 (⇊) | 0.76 | 0.15 | 4.42 |
| 89 | Govind Ballabh Pant University of Agriculture Technology | 1,344 | 89 | 6.78 | 0.3 | 2,270 | 162 (⇊) | 9.63 | 0.19 | 6.32 |
| 90 | Pondicherry University | 1,319 | 90 | 19.05 | 0.3 | 4,826 | 53 (↑) | 7.96 | 0.4 | 13.96 |
| 91 | Maulana Azad Medical College | 1,317 | 91 | 14.77 | 0.3 | 2,223 | 167 (⇊) | 3.63 | 0.18 | 5.72 |
| 92 | Bose Institute | 1,307 | 92 | 8.81 | 0.29 | 2,286 | 161 (⇊) | 1.08 | 0.19 | 4.93 |
| 93 | Kasturba Medical College Manipal | 1,275 | 93 | 18.43 | 0.29 | 3,945 | 77 (↑) | 12.47 | 0.33 | 15.04 |
| 94 | Birla Institute of Technology Mesra | 1,264 | 94 | 24.78 | 0.28 | 4,036 | 74 (↑) | 12.64 | 0.33 | 18.58 |
| 95 | National Institute of Pharmaceutical Education Research SAS Nagar Mohali | 1,186 | 95 | 6.22 | 0.27 | 1,481 | 257 (⇊) | -0.12 | 0.12 | 3.37 |
| 96 | Dr Hari Singh Gour University | 1,173 | 96 | 7 | 0.26 | 1,481 | 256 (⇊) | 5.29 | 0.12 | 4.84 |
| 97 | ICAR-National Dairy Research Institute | 1,148 | 97 | 13.71 | 0.26 | 3,397 | 101 (⇊) | 4.56 | 0.28 | 9.79 |
| 98 | Tezpur University | 1,148 | 98 | 31.2 | 0.26 | 3,741 | 85 (↑) | 4.91 | 0.31 | 18.16 |
| 99 | Institute of Physics Bhubaneswar (IOPB) | 1,111 | 99 | -2.32 | 0.25 | 2,336 | 155 (⇊) | 6.56 | 0.19 | 3.63 |
| 100 | CSIR-Central Glass Ceramic Research Institute CGCRI | 1,107 | 100 | 10.57 | 0.25 | 1,749 | 220 (⇊) | -2.43 | 0.15 | 4.38 |
| 101 | Alagappa University | 1,092 | 103 | 14.4 | 0.25 | 3,488 | 97 (⇈) | 9.89 | 0.29 | 10.63 |
| 102 | National Institute of Technology (NIT) Karnataka | 1,018 | 112 | 25.84 | 0.23 | 4,904 | 50 (⇈) | 16.65 | 0.41 | 19.09 |
| 103 | SRM Institute of Science Technology Chennai | 987 | 119 | 69.15 | 0.22 | 12,742 | 17 (⇈) | 28.38 | 1.06 | 43.36 |
| 104 | Indian Institute of Technology (IIT) Indian School of Mines Dhanbad | 982 | 120 | 26.16 | 0.22 | 10,135 | 23 (⇈) | 18.02 | 0.84 | 21.58 |
| 105 | Jawaharlal Institute of Postgraduate Medical Education Research | 841 | 140 | 23.16 | 0.19 | 3,818 | 82 (⇈) | 10.30 | 0.32 | 14.01 |
| 106 | National Institute of Technology (NIT) Durgapur | 835 | 141 | 27.81 | 0.19 | 3,864 | 80 (⇈) | 13.87 | 0.32 | 18.99 |
| 107 | University Of Kashmir | 783 | 149 | 28.91 | 0.18 | 4,323 | 69 (⇈) | 15.70 | 0.36 | 20.71 |
| 108 | Indian Institute Of Science Education Research (IISER) Kolkata | 742 | 163 | 66.06 | 0.17 | 4,085 | 72 (⇈) | 9.25 | 0.34 | 34.40 |
| 109 | Shanmugha Arts Science Technology Research Academy (SASTRA) | 728 | 167 | 40.58 | 0.16 | 5,683 | 41 (⇈) | 8.75 | 0.47 | 25.40 |
| 110 | Motilal Nehru National Institute of Technology (NIT) | 707 | 171 | 26.54 | 0.16 | 3,431 | 98 (⇈) | 14.81 | 0.28 | 17.86 |
| 111 | Sardar Vallabhbhai National Institute of Technology (NIT) | 703 | 173 | 65.67 | 0.16 | 3,804 | 83 (⇈) | 12.81 | 0.32 | 34.77 |
| 112 | National Institute of Technology (NIT) Warangal | 679 | 181 | 17.76 | 0.15 | 4,749 | 57 (⇈) | 21.05 | 0.39 | 18.64 |
| 113 | National Institute of Technology (NIT) Calicut | 624 | 193 | 37.55 | 0.14 | 3,407 | 100 (⇈) | 20.16 | 0.28 | 25.10 |
| 114 | Delhi Technological University | 610 | 197 | 27.56 | 0.14 | 4,786 | 55 (⇈) | 25.67 | 0.4 | 23.70 |
| 115 | Indian Institute of Science Education Research (IISER) Pune | 536 | 213 | 64.46 | 0.12 | 4,487 | 61 (⇈) | 11.17 | 0.37 | 34.12 |

| | | | | | | | | | |
|---|---|---|---|---|---|---|---|---|---|
| 116 | Siksha O Anusandhan University | 501 | 222 | 78.93 | 0.11 | 4,929 | 49 (⇈) | 19.60 | 0.41 | 41.54 |
| 117 | Gandhi Institute Of Technology Management (GITAM) | 500 | 223 | 27.78 | 0.11 | 4,082 | 73 (⇈) | 29.40 | 0.34 | 23.97 |
| 118 | Amity University NOIDA | 491 | 227 | 73.84 | 0.11 | 6,873 | 33 (⇈) | 26.16 | 0.57 | 43.30 |
| 119 | Visvesvaraya National Institute Of Technology (NIT) Nagpur | 433 | 243 | 25.64 | 0.10 | 3,529 | 95 (⇈) | 21.12 | 0.29 | 23.41 |
| 120 | SSN College of Engineering | 431 | 244 | 48.68 | 0.10 | 3,561 | 94 (⇈) | 21.47 | 0.3 | 35.27 |
| 121 | Malaviya National Institute of Technology (NIT) Jaipur | 415 | 252 | 26.39 | 0.09 | 4,384 | 67 (⇈) | 26.44 | 0.36 | 23.59 |
| 122 | Indian Institute of Technology (IIT) Hyderabad | 377 | 266 | 74.5 | 0.09 | 6,071 | 37 (⇈) | 19.61 | 0.5 | 43.94 |
| 123 | Sathyabama Institute of Science Technology | 337 | 282 | 31.8 | 0.08 | 4,253 | 70 (⇈) | 25.33 | 0.35 | 27.63 |
| 124 | Datta Meghe Institute of Higher Education Research (Deemed to be University) | 329 | 288 | 58.74 | 0.07 | 4,763 | 56 (⇈) | 29.26 | 0.4 | 29.25 |
| 125 | Kalinga Institute of Industrial Technology (KIIT) | 315 | 296 | 70.69 | 0.07 | 5,141 | 45 (⇈) | 22.78 | 0.43 | 41.66 |
| 126 | Indian Institute Of Technology (IIT) Bhubaneswar | 273 | 333 | 71.74 | 0.06 | 3,613 | 90 (⇈) | 16.77 | 0.3 | 39.21 |
| 127 | Indian Institute of Technology (IIT) Patna | 230 | 387 | 65.07 | 0.05 | 3,625 | 89 (⇈) | 24.06 | 0.3 | 41.13 |
| 128 | Indian Institute of Technology (IIT) Indore | 222 | 397 | 44.22 | 0.05 | 5,741 | 40 (⇈) | 19.60 | 0.48 | 31.81 |
| 129 | Saveetha Institute of Medical Technical Science | 198 | 420 | 57.01 | 0.05 | 10,777 | 20 (⇈) | 61.02 | 0.89 | 53.29 |
| 130 | Lovely Professional University | 187 | 442 | 32.69 | 0.04 | 6,481 | 35 (⇈) | 43.57 | 0.54 | 36.75 |
| 131 | Koneru Lakshmaiah Education Foundation K L (Deemed to be University) | 180 | 461 | 56.4 | 0.04 | 4,447 | 64 (⇈) | 42.40 | 0.37 | 43.56 |
| 132 | National Institute of Technology (NIT) Silchar | 168 | 480 | 26.34 | 0.04 | 3,926 | 78 (⇈) | 32.31 | 0.33 | 30.25 |
| 133 | Homi Bhabha National Institute | 152 | 519 | 38.7 | 0.03 | 14,993 | 11 (⇈) | 48.56 | 1.24 | 41.82 |
| 134 | University Of Petroleum Energy Studies (UPES) | 84 | 752 | 49.81 | 0.02 | 4,458 | 63 (⇈) | 44.99 | 0.37 | 46.03 |
| 135 | Saveetha Dental College Hospital | 78 | 788 | 43 | 0.02 | 5,510 | 43 (⇈) | 65.65 | 0.46 | 45.93 |
| 136 | GLA University | 76 | 806 | 22.76 | 0.02 | 3,645 | 88 (⇈) | 70.38 | 0.3 | 37.82 |
| 137 | Symbiosis International University | 49 | 1,092 | 25.99 | 0.01 | 3,504 | 96 (⇈) | 48.39 | 0.29 | 32.08 |
| 138 | Chandigarh University | - | - | 0 | - | 4,440 | 65 (⇈) | 97.99 | 0.37 | 48.66 |
| 139 | ICAR-Indian Agricultural Research Institute | - | - | 11.03 | - | 7,393 | 31 (⇈) | 10.45 | 0.61 | 9.88 |
| 140 | Saveetha School Of Engineering | - | - | 8.01 | - | 3,852 | 81 (⇈) | 65.91 | 0.32 | 37.57 |

In between the two periods a total of 40 institutions moved out of the top 100 list based on their publication output. Among them eighteen (18) are universities (16 state and 2 central), eleven (11) are a part of the Council for Scientific and Industrial Research (CSIR), while others are different other institutions (Table 2). This may be related with findings reported in an earlier study, where it was found that CSIR system's proportionate share in national research output has declined from 2001-2005 to 2016-2020 (Kanaujia et. al. 2022).

Now we look at individual institutions which were in the top 100 list in the 1st period, but went out of the list in the 2nd time period. The major Universities moving out of top 100 list were, University of Mysore which moved from 29th to 113th position, University of Rajasthan which moved from 35th to 191st position, University of Allahabad which moved from 41st to 104th position, Sri Venkateswara University which moved from 44th to 142nd position, Maharaja Sayajirao University Baroda which moved from 47th to 111th position, Mangalore University which moved from 48th to 193rd position, Andhra University which moved from 49th to 106th position, Shivaji University which moved from 55th to 120th position, University of Mumbai which moved from 57th to 172nd position, Lucknow University which moved from 61st to 123th position, Madurai Kamaraj University which moved from 62nd to 129th position, Kalyani University which moved from 65th to 117th position, Karnataka University which moved from 68th to 179th position, Kurukshetra University which moved from 74th to 149th position, Bangalore University which moved from 77th to 218th position, Jawaharlal Nehru Technological University Hyderabad which moved from 78th to 146th position, Govind Ballabh Pant University Of Agriculture Technology which moved from 89th to 162th position, and Dr Hari Singh Gour University which moved from 96th to 256th position. Thus, many state Universities which were among the top 100 contributors to Indian research output during 2004-13 moved out of this list during 2014-23.

Several CSIR institutions have also lost their position, such as Central Drug Research Institute (CDRI) moved from 33rd to 107th position, Central Food Technological Research Institute (CFTRI) moved from 53rd to 176th position, National Institute Interdisciplinary Science Technology (NIIST) from 52nd to 139th position, Central Leather Research Institute (CLRI) moved from 59th to 185th position, Indian Institute Of Chemical Biology (IICB) moved from 70th to 194th position, National Institute of Oceanography (NIO) moved from 73th to 168th position, Central Electrochemical Research Institute (CECRI) moved from 76th to 138th position, Central Salt Marine Chemical Research Institute (CSMCRI) 82nd to 156th position, Centre For Cellular Molecular Biology (CCMB) moved from 86th to 255th position, National Geophysical Research Institute (NGRI) moved from 88th to 206th position, Central Glass Ceramic Research Institute (CGCRI) moved from 100th to 220th position. This shift may be either due to decrease in research activities and hence publications of these institutions or possibly due to a change of nature of research outcomes from these institutes (say patents, technology transfers etc. instead of research papers).

Among the other institutions, Jawaharlal Nehru Center for Advanced Scientific Research (JNCASR) moved from 34th to 110th position, Physical Research Laboratory (PRL) India moved from 75th to 134th position, Inter University Accelerator Centre moved from 79th to 203rd position, UGC DAE Consortium for Scientific Research moved from 81st to 128th position, SN Bose National Centre for Basic Science (SNBNCBS) moved from 84th to 163rd position, Raja Ramanna Centre for Advanced Technology moved from 87th to 196th position, Maulana Azad Medical College moved from 91st to 167th position, Bose Institute moved from 92nd to 161st position, National Institute of Pharmaceutical Education Research (NIPER) S A S Nagar Mohali moved from 95th to 257th position, ICAR National Dairy Research Institute moved from 97th to 101st position, and Institute Of Physics Bhubaneswar (IOPB) moved from 99th to 155th position. Taken together, the contribution of these 40 organisations in national output during the 1st period was 14.7% of publication in the first period, which reduced to 8.01% during the 2nd period.

Other major organisations whose position in terms of percentage of publications reduced were CSIR-Indian Institute of Chemical Technology (IICT), Kolkata (13 to 38), CSIR-National Chemical Laboratory (NCL), Pune (19 to 46), Indian Association for the Cultivation Of Science (IACS), Jadavpur (20 to 60), Annamalai University (21 to 42), University Of Madras

(22 to 52), and CSIR-National Physical Laboratory (NPL), Delhi (27 to 87). The highest CAGR over both periods among these institutions is 10.24% for Jawaharlal Nehru Technical University Hyderabad. From table 1 it may also be noted that many of these institutions had a negative CAGR in the 2nd period indicating a slower rate of publication during the period. The movement of central and state universities, central government funded institutions (Bose Institute, IOPB, JNCASR, PRL, Raja Ramanna Centre for Advanced Technology, SNBNCBS), CSIR, ICAR, NIPER Institutes and UGC centres etc. out of the top 100 institutions list may be an indication towards a change in research agenda with higher focus on other activities such as technology development, patents, etc. in which case, there may be a need to explore possible indicators which can capture these activities. It may also be an indication towards a possible shortcoming in their research agenda/policies. Either these institutions have not evolved their research policies to promote higher output from their research activities or they have not received the necessary attention and resources to scale up and sustain the growth in their research activities.

**Table 2: Institutions which went out of top 100 list in 2014-23 period**

| S. No. | Institution name | Type |
|---|---|---|
| 1 | Andhra University | State |
| 2 | Bangalore University | State |
| 3 | Bose Institute | DST |
| 4 | CSIR-Central Drug Research Institute (CDRI) | CSIR |
| 5 | CSIR-Central Electrochemical Research Institute (CECRI) | CSIR |
| 6 | CSIR-Central Food Technological Research Institute (CFTRI) | CSIR |
| 7 | CSIR-Central Glass Ceramic Research Institute (CGCRI) | CSIR |
| 8 | CSIR-Central Leather Research Institute (CLRI) | CSIR |
| 9 | CSIR-Central Salt Marine Chemical Research Institute (CSMCRI) | CSIR |
| 10 | CSIR-Centre For Cellular Molecular Biology (CCMB) | CSIR |
| 11 | CSIR-Indian Institute Of Chemical Biology (IICB) | CSIR |
| 12 | CSIR-National Geophysical Research Institute (NGRI) | CSIR |
| 13 | CSIR-National Institute Interdisciplinary Science Technology (NIIST) | CSIR |
| 14 | CSIR-National Institute of Oceanography (NIO) | CSIR |
| 15 | Dr Hari Singh Gour University | Central |
| 16 | Govind Ballabh Pant University of Agriculture Technology | State |
| 17 | ICAR-National Dairy Research Institute (NDRI) | ICAR |
| 18 | Institute of Physics Bhubaneswar (IOPB) | DAE |
| 19 | Inter University Accelerator Centre | UGC |
| 20 | Jawaharlal Nehru Center for Advanced Scientific Research (JNCASR) | DST |
| 21 | Jawaharlal Nehru Technological University, Hyderabad | State |
| 22 | Kalyani University | State |
| 23 | Karnatak University | State |
| 24 | Kurukshetra University | State |
| 25 | Lucknow University (University of Lucknow) | State |
| 26 | Madurai Kamaraj University | State |
| 27 | Maharaja Sayajirao University Baroda | State |
| 28 | Mangalore University | State |
| 29 | Maulana Azad Medical College | State |
| 30 | National Institute of Pharmaceutical Education Research, SAS Nagar Mohali | NIPER |

| 31 | Physical Research Laboratory India | DoS |
|---|---|---|
| 32 | Raja Ramanna Centre for Advanced Technology | DAE |
| 33 | Shivaji University | State |
| 34 | Sn Bose National Centre for Basic Science Snbncbs | DST |
| 35 | Sri Venkateswara University | State |
| 36 | UGC DAE Consortium for Scientific Research | UGC |
| 37 | University of Allahabad | Central |
| 38 | University of Mumbai | State |
| 39 | University of Mysore | State |
| 40 | University of Rajasthan | State |

New entrants in the 2013-2024 period are mainly Private universities (18), NITs (9), IITs (5), IISERs (2) which were newly established (Table 3). Three (3) state universities, Alagappa University (97), Delhi Technical University (55) and University of Kashmir (69) also entered the top-100 institutions. Among the private universities major changes can be observed in the total number of publications from having only a few hundred publications to thousands of publications indicating a major change in publication trends in these institutions. These are, (1) SRM Institute of Science Technology Chennai 119th to 17th position, (2) Shanmugha Arts Science Technology Research Academy SASTRA 170th to 40th position, (3) Sardar Vallabhbhai National Institute of Technology from 173rd to 83rd position, (4) Siksha O Anusandhan University from 222th to 49th position, (5) Gandhi Institute of Technology Management GITAM from 223rd to 73rd position. (6) Amity University NOIDA from 227th to 33rd position, (7) SSN College of Engineering from 244th to 94th position, (8) Sathyabama Institute of Science Technology from 282nd to 70th position, (9) Datta Meghe Institute of Higher Education Research Deemed To Be University from 288th to 56th position, (10) Kalinga Institute of Industrial Technology KIIT from 296th to 45th position, (11) Lovely Professional University from 442nd to 35th position, (12) Koneru Lakshmaiah Education Foundation K L Deemed To Be University from 461st to 64th position, (13) Saveetha Dental College Hospital from 788th to 43rd position, (14) GLA University from 806th to 88th position, (15) Symbiosis International University from 1092nd to 96th position, (16) Chandigarh University was established in 2012 and it is placed at 65th position in the 2014-2023 period. A closer look at the newly entering private universities suggests that many of them are large universities with more than 500 faculty members (QS Top University Profiles-https://www.topuniversities.com/universities/).

Among the NITs, the nine institutes that gained are Malaviya NIT Jaipur (252 to 67), Motilal Nehru NIT in Prayagraj (171 to 98), NIT Calicut (193 to 100), NIT Durgapur (141 to 80), NIT Karnataka (112 to 50), NIT Silchar (480 to 78), NIT Warangal (181 to 57), Sardar Vallabhbhai NIT (173 to 83), and Visvesvaraya NIT Nagpur (243 to 95). Five IITs are four newly established institutes namely IIT Bhubaneswar (333 to 90), IIT Hyderabad (266 to 37), IIT Indore (397 to 40), IIT Patna (387 to 89), and IIT-ISM Dhanbad (120 to 23) entered the top 100 list in the second period. JIPMER (140 to 82) and two IISERs Pune (213 to 62) and Kolkata (163 to 72) also entered the list. These 40 institutions taken together account for 17.57% of the total research publications in the 2nd period.

**Table 3: Institutions which entered the top 100 list in 2014-23 period**

| S. No. | Institution Name | Type |
|---|---|---|
| 1 | Alagappa University | State |
| 2 | Amity University NOIDA | Private |
| 3 | Chandigarh University | Private |
| 4 | Datta Meghe Institute of Higher Education Research (Deemed To Be University) | Private |
| 5 | Delhi Technological University (DTU) | State |
| 6 | Gandhi Institute of Technology Management (GITAM) | Private |
| 7 | GLA University | Private |
| 8 | Homi Bhabha National Institute | DAE |
| 9 | ICAR-Indian Agricultural Research Institute | ICAR |
| 10 | Indian Institute of Science Education Research (IISER) Kolkata | IISER |
| 11 | Indian Institute of Science Education Research (IISER) Pune | IISER |
| 12 | Indian Institute of Technology (IIT) Bhubaneswar | IIT |
| 13 | Indian Institute of Technology (IIT) Hyderabad | IIT |
| 14 | Indian Institute of Technology (IIT) Indore | IIT |
| 15 | Indian Institute of Technology (IIT) Patna | IIT |
| 16 | Indian Institute of Technology (IIT) Indian School Of Mines Dhanbad | IIT |
| 17 | Jawaharlal Institute of Postgraduate Medical Education Research | JIPMER |
| 18 | Kalinga Institute of Industrial Technology (KIIT) | Private |
| 19 | Koneru Lakshmaiah Education Foundation K L (Deemed To Be University) | Private |
| 20 | Lovely Professional University | Private |
| 21 | Malaviya National Institute of Technology (NIT) Jaipur | NIT |
| 22 | Motilal Nehru National Institute Of Technology (NIT) | NIT |
| 23 | National Institute of Technology (NIT) Calicut | NIT |
| 24 | National Institute of Technology (NIT) Durgapur | NIT |
| 25 | National Institute of Technology (NIT) Karnataka | NIT |
| 26 | National Institute of Technology (NIT) Silchar | NIT |
| 27 | National Institute of Technology (NIT) Warangal | NIT |
| 28 | Sardar Vallabhbhai National Institute of Technology (NIT) | NIT |
| 29 | Sathyabama Institute of Science Technology | Private |
| 30 | Saveetha Dental College Hospital | Private |
| 31 | Saveetha Institute of Medical Technical Science | Private |
| 32 | Saveetha School of Engineering | Private |
| 33 | Shanmugha Arts Science Technology Research Academy (SASTRA) | Private |
| 34 | Siksha O Anusandhan University | Private |
| 35 | SRM Institute of Science Technology Chennai | Private |
| 36 | SSN College of Engineering | Private |
| 37 | Symbiosis International University | Private |
| 38 | University of Kashmir | State |
| 39 | University of Petroleum Energy Studies (UPES) | Private |
| 40 | Visvesvaraya National Institute of Technology (NIT) Nagpur | NIT |

## 5. Discussion

The research publications from Indian Institutions during two periods 2004-2013 and 2014-2023 were analysed and the major contribution institutions during the two periods were identified. The total publications (TP), share in national TP and the CAGR of publications of the top 100 institutions were computed and their respective management and funding types were found out and presented. Top 3 institutions remained the same during both periods, namely, IISc, IIT Kharagpur and University of Delhi. In the Top 10 list, 2 institutions in first period IIT Kanpur (9) and Jadavpur University (10) moved out while IIT Roorkee (10) and VIT University were placed at 10th and 8th positions respectively in the second period. In the top 100 list of 2004-13, there are eight (8) IITs, eleven (11) central universities, thirty-two (32) state universities, fourteen (14) CSIR labs, seven (7) private universities and 28 others. During 2014-23 period, this list changes to include a total of thirteen (13) IITs, nine (9) central universities, nineteen (19) state universities, three (3) CSIR labs, twenty-four (24) private universities and 32 others. **(RQ1)**.

A total of 40 institutions were replaced in the second period by other institutions. Among the institutions which lost their position there were eighteen (18) state universities, and eleven (11) CSIR labs. In the second time period several private universities (18), NITs (9), IITs (5), IISERs (2), state universities (3), and some centrally funded institutions entered the top-100 institutions list **(RQ2)**. It is to be noted that, in 2003, the government introduced regulation for establishment and maintenance of standards in private universities (UGC 2003), and in 2008 the recommendations of National Knowledge Commission (NKC) report "Towards a Knowledge Society: Three Years of the National Knowledge Commission" were released which also made the case for increasing private investment in education. These paved the way for establishment of new large private universities which are multidisciplinary and focus on research in addition to regular education activities (Sengupta, 2020).

It is observed that several government-funded institutions have been overtaken by private universities and Institutions of national importance having express focus on research along with good funding support. Several of the new entering institutes in the top 100 list during the 2nd period are private universities, institutes of national importance, such as newly formed IITs and IISERs. The NITs which were formed from the Regional Engineering Centres in the early 2000s have also moved into the list with a large shift in their publication outputs. Three state universities Alagappa, DTU and University of Kashmir and a handful of central government/department funded institutes were added to the top 100 list in the 2nd period. In this regard, it may be noted that DTU is an old institution established in 1941 as a polytechnic, and started awarding degrees under the University of Delhi in 1952 and given university status in 2009, Alagappa University was established in 1985 and University of Kashmir was established in 1948. The University of Kashmir saw a major infrastructural development with the establishment of its Satellite Campuses in 2008 onwards which could be a contributing factor in increasing the research publication output of the university.

The results also indicate that several state universities and CSIR labs have moved out of the top 100 list during the 2nd time period. In most cases the total number of publications from them did not increase much during the period. The low growth in number of publications of universities may indicate increasing problems in the research ecosystem in these institutions and possibly also a sign of the research infrastructure becoming outdated. In case of CSIR, a major systemic change happened with the establishment of Academy of Scientific and Innovative Research (AcSIR), a national university with CSIR labs as its campus hubs and scientists as professors. This injected a large number of new research students in the CSIR system and may have led to changes in productivity profiles of the institutes. However, despite this increase in number of researchers, the growth of number of publications from various CSIR labs have been low. The slower growth in number of publications from CSIR labs may be due to a change in focus on activities other than research publication. These may include, technology development, incubation, technology transfer, patents, educational programmes, strategic international partnerships etc.

When the patterns of growth in number of publications of certain institutions is seen along with the increase in the number of Institutes of National Importance and private universities; one may like to infer that the several newly formed institutes have started contributing to the national research output in a short span of time and possibly some more such institutions may enter in this list in the next few years. The several private institutions which have entered the top 100 indicate a significant shift in the outlook and the role of the private sector in education and research in India. The changing patterns in case of the CSIR system may also be examined more closely to ascertain whether this is due to change in activities of the labs or that it is a sign of slowdown of research in these labs. A good number of state universities going out of the top 100 list is also a disturbing observation that needs a serious and detailed examination.

## 5. Conclusion

This study looked at the research output from Indian universities, and research institutions across two periods 2004-13 and 2014-23 and identified the ones which had highest number of publications. It looked for changes in the composition of the top 100 list tried to identify the type of institutions which entered the top 100 list in 2$^{nd}$ period and also those institutions which moved out of the top 100 list in the 2$^{nd}$ period. The most notable observations are in terms of the growth in number of private universities and newly establishment INIs which have made their entry in the in top 100 list. These institutions replaced several state universities, and centrally funded research institutions such as the CSIR laboratories, ICAR and NIPER institutes. Thus, these observations indicate towards a possible institutional shift in contribution to the research publications of India. The state universities and certain CSIR labs going out of the top 100 list is a disturbing observation that needs to examined in more detail to concretely identify the probable reasons.

We would like to also state here that the findings of this study are based on the research publications data listed on the Web of Science database. As a result, the inferences drawn here may not have accounted for research productivity of institutions and universities coming out in form of patents, conferences, workshops, bilateral or multilateral collaborations, technology developments, incubation, and transfer activities etc. Additionally, any publication in journals not listed in the Web of Science database are not accounted for. As it is difficult to get information about several of these, it is a possible area that can be explored in future studies. This can possibly be done with support of the funding agencies and the universities and institutions themselves.